\documentstyle[prb,aps,epsf,eqsecnum,umlaut]{revtex} 
\begin{document}
\draft
\twocolumn[\hsize\textwidth\columnwidth\hsize\csname@twocolumnfalse%
\endcsname
\title{Impurity spin relaxation in $S=1/2$ $XX$ chains}
\author{Joachim Stolze and Michael Vogel} 
\address{Institut f\"ur Physik, Universit{\"a}t
  Dortmund, 44221 Dortmund, Germany } 
\date{\today}
\maketitle
\begin{abstract}
  Dynamic autocorrelations $\langle S_i^{\alpha} (t) S_i^{\alpha} \rangle$
  ($\alpha=x,z$) of an isolated impurity spin in a $S=1/2$ $XX$ chain are
  calculated. The impurity spin, defined by a local change in the
  nearest-neighbor coupling, is either in the bulk or at the boundary of the
  open-ended chain. The exact numerical calculation of the correlations
  employs the Jordan-Wigner mapping from spin operators to Fermi operators;
  effects of finite system size can be eliminated. Two distinct temperature
  regimes are observed in the long-time asymptotic behavior of the bulk
  correlations.  At $T=0$ only
  power laws are present. At high $T$ the $x$ correlation decays exponentially
  (except at short times) while the $z$ correlation still shows an asymptotic
  power law (different from the one at $T=0$) after an intermediate
  exponential phase. The boundary impurity correlations ultimately follow
  power laws at 
  all $T$, with intermediate exponential phases at high $T$.  The power laws
  for the $z$ correlation and the boundary 
  correlations can be derived from the impurity-induced changes in the
  properties of the Jordan-Wigner fermion states.
\end{abstract}
\vskip 0.03 truein
]
\section{Introduction}
\label{I}
The $S=1/2$ $XX$ chain \cite{LSM61,Kat62} 
\begin{equation}
  \label{2.1}
  H=- \sum_{i=1}^{N-1} J_i (S_i^x S_{i+1}^x +S_i^y S_{i+1}^y) ,
\end{equation}
is one of the simplest quantum many-body systems conceivable, as many of its
properties can be derived from those of noninteracting lattice fermions. Its
equilibrium spin pair correlation functions
\begin{equation}
  \label{2.6}
  \langle S_i^{\alpha}(t) S_j^{\alpha} \rangle
= \frac
{{\rm Tr} e^{-\beta H} e^{itH} S_i^{\alpha} e^{-itH} S_j^{\alpha}}{{\rm Tr}
  e^{-\beta H}} , \alpha =x,z
\end{equation}
have been the objects of intense research efforts over an extended
period
\cite{Nie67,KHS70,Tjo70,MBA71,SJL75,BJ76,CP77,PM78,VT78,GC80,CG81,MPS83a,MPS83b,MS84,TM85,FL87,my18,IIKS93,my28}.
Only a few explicit analytic results are available, but several existing
 asymptotic results for large distances $|i-j|$ or long times $t$ have been
 corroborated by numerical calculations. 
 
 For $XX$ chains with homogeneous nearest-neighbor coupling ($J_i \equiv J$)
 only three different types of asymptotic long-time behavior have been
 observed to date: Gaussian, exponential, and power-law (often with
 superimposed oscillations).  It is interesting to speculate whether
 non-uniform or random couplings might induce additional types of asymptotic
 behavior.
 
 In the present paper we study the changes in autocorrelation functions
 $\langle S_i^{\alpha}(t) S_i^{\alpha} \rangle$ induced by a single impurity
 spin in an otherwise homogeneous chain. The impurity spin is located either
 at the boundary of the system,
\begin{equation}
  \label{2.7}
  J_1=J^{\prime} , J_i \equiv J =1 \mbox{ for } i \geq 2
\end{equation}
or in the bulk,
\begin{equation}
  \label{2.8}
  J_{N/2-1} = J_{N/2} = J^{\prime} , J_i \equiv J =1 \mbox{ for all other } i.
\end{equation}

Equilibrium and non-equilibrium dynamics of the boundary impurity were studied
early on by Tjon \cite{Tjo70}. In the weak-coupling limit ($J^{\prime}
\rightarrow 0$), Tjon obtained exponential behavior of the impurity spin
autocorrelation functions. Our results (see Sec. \ref{IV}) show that for
finite impurity coupling $J^{\prime}$, exponential behavior occurs only in an
intermediate time regime, whereas the ultimate long-time behavior is a power law.

Besides their obvious relevance in low-dimensional magnetism, impurities in
spin-1/2 chains are also of interest in quantum dynamics, where two-level
systems coupled to ``baths'' serve as models for quantum systems in dissipative
environments \cite{LCDFGZ87,Wei93,HTB90}.  The most popular model \cite{Wei93}
in this field is the spin-boson model, consisting of a single spin 1/2 coupled
to a (quasi-) continuum of noninteracting oscillators with a given spectral
density.  In a recent study\cite{SH98} the oscillator bath was replaced with a
bath of noninteracting spins 1/2. The changes in dynamic behavior which were
observed as a result of this replacement suggest further exploration of
different kinds of baths.  The system studied here can be considered a
two-level system (the impurity spin) coupled to a bath of {\em interacting}
two-level systems (the remainder of the $XX$ chain). An interesting feature of
this system is the fact that while the $z$ component of the total spin is
conserved, the $x$ component is not. Thus differences are to be expected
between the relaxation of the $x$ and $z$ components of the impurity spin.
 
The plan of the paper is as follows: In Sec. \ref{II} we discuss the method
used to calculate the dynamic correlation functions numerically. In Sec.
\ref{III} we present results for spin autocorrelation functions of a bulk
impurity spin (and also of its neighbors) for both zero and finite $T$. In
Sec. \ref{IV} we discuss boundary impurity autocorrelation functions. Sec.
\ref{V} summarizes our findings.
\section{Method}
\label{II}
The open-ended $N$-site spin-1/2 $XX$ chain described by the Hamiltonian
(\ref{2.1}) can be mapped to a Hamiltonian of noninteracting fermions,
\begin{equation}
  \label{2.2}
  \tilde{H}= - \frac{1}{2} \sum_{i=1}^{N-1} J_i (c_i^{\dag} c_{i+1} +
  c_{i+1}^{\dag} c_i)
\end{equation}
by means of the Jordan-Wigner transformation \cite{LSM61,Kat62} between spin
and fermion operators:
\begin{equation}
  \label{2.3}
  S_i^z = c_i^{\dag} c_i - \frac{1}{2},
\end{equation}
\begin{equation}
  \label{2.4}
  S_i^+ = (-1)^{\sum_{k=1}^{i-1}c_k^{\dag} c_k} c_i^{\dag} = \prod_{k=1}^{i-1}
  (1 - 2 c_k^{\dag} c_k) c_i^{\dag}.
\end{equation}
In the homogeneous case $J_i \equiv J$, the one-particle energy eigenvalues
are
\begin{equation}
  \label{2.5}
  \varepsilon_k = -J \cos k , k= \frac{\nu \pi}{N+1}
, \nu= 1, \cdots,N
\end{equation}
and the eigenvectors are sinusoidal functions of the site index $i$. For
general $J_i$ neither eigenvalues nor eigenvectors are available analytically,
however, both are easily obtained from the solution of a tridiagonal
eigenvalue problem with standard numerical procedures \cite{PTVF92}.

The spin correlation functions (\ref{2.6}) are mapped to fermion correlation
functions, with crucial differences between the cases $\alpha=z$ and
$\alpha=x$. $\langle S_i^{z}(t) S_j^{z} \rangle$ maps to a density-density
correlation function involving four Fermi operators. Due to the string of
signs in (\ref{2.4}), however, $\langle S_i^{x}(t) S_j^{x} \rangle$ maps to a
{\em many-particle} correlation function involving $2(i+j-1)$ Fermi operators.
Wick's theorem can be applied to expand $\langle S_i^{x}(t) S_j^{x} \rangle$
in products of elementary fermion expectation values. That expansion can be
most compactly expressed as a Pfaffian \cite{Pfaffnote} whose elements are
sums involving the one-particle eigenvalues and eigenvectors.

In order to obtain results valid in the thermodynamic limit $N \rightarrow
\infty$, finite-size effects must be identified and eliminated. As finite-size
effects are known \cite{my28} to be caused by reflections of propagating
excitations from the boundaries of the system, the maximum fermion group
velocity (see \ref{2.5}) can be used to estimate the time range over which a
given spin correlation function (\ref{2.6}) can be expected to be free of
finite-size effects. That estimate can then be verified by explicit numerical
calculation of equivalent correlation functions for system sizes $N_0$ and,
say, $2N_0$.

The one-fermion eigenvalue problem for the single-impurity chain (\ref{2.7})
or (\ref{2.8}) \cite{note2} may be solved analytically. The nature of the
solution depends on the value of $J^{\prime}$. For $J^{\prime}$ below a
critical value $J_c$ all states are extended and the continuous energy
spectrum is given by $\varepsilon_k$ (\ref{2.5}) with $J^{\prime}$-dependent
$k$ values. For $J^{\prime} > J_c$ a pair of exponentially localized impurity
states with energies $\pm \varepsilon_0$, $|\varepsilon_0| > 1$, emerge from
the continuum. The critical coupling strength is $J_c=1$ for the bulk impurity
and $J_c = \sqrt{2}$ for the boundary impurity.  Below, we shall occasionally
refer to properties of the analytic solution in order to explain the long-time
asymptotic behavior observed in the numerical results.
\section{Bulk impurity}
\label{III}
\subsection{$T=0$}
\label{IIIA}
%%%%%%%%%%%%%%%%%%%%%%%%%%%%%%%%%%%%%%%%%%%%%%%%%%%%%%%%%%%%%%%%%%%%
\begin{figure}[h] 
\centerline{ \epsfxsize=9cm \epsfbox{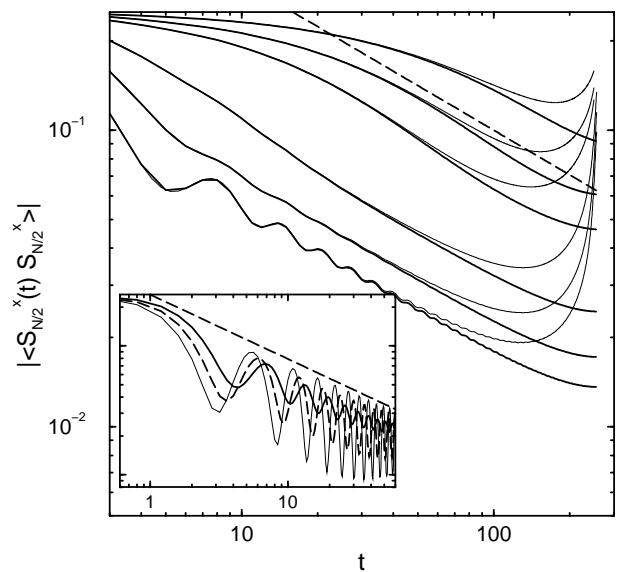} } 
\caption{Bulk impurity spin $x$ autocorrelation function 
$|\langle S_{N/2}^x(t) S_{N/2}^x \rangle|$ at $T=0$.
 Main plot: $N=512$, impurity coupling $J^{\prime}<1$. The heavy solid lines
 (top to bottom) correspond to $J^{\prime}=0.1$, 0.15, 0.2, 0.4, 0.6, and 0.8.
 The dashed straight line is the $t^{-1/2}$ power law (\ref{3.1}). The thin
 solid lines are $N=256$ data demonstrating the influence of system size.
Inset: $N=128$,  $J^{\prime}>1$. Shown are data for $J^{\prime}=1$ (heavy
 solid line), $J^{\prime}=1.2$ (Dashed curve), and $J^{\prime}=1.4$ (thin
 solid line). The dashed straight line is again the $t^{-1/2}$ power law
 (\ref{3.1}). }
\label{eins} 
\end{figure}
%%%%%%%%%%%%%%%%%%%%%%%%%%%%%%%%%%%%%%%%%%%%%%%%%%%%%%%%%%%%%%%
The long-time asymptotic behavior of the $T=0$ bulk impurity spin $x$
autocorrelation function is difficult to obtain due to a combination of two
reasons. Firstly this correlation is the computationally most demanding one,
as large Pfaffians have to be evaluated. Secondly it is also the correlation
function displaying finite size effects at the earliest times.  This may be
related to its particularly slow long-time asymptotic decay law
\cite{VT78,MPS83a,MPS83b}
\begin{equation}
  \label{3.1}
  \langle S_i^x(t)S_{i+n}^x\rangle \sim (n^2 - J^2 t^2)^{-1/4} \mbox{ for }T=0
\end{equation}
in the homogeneous case $J_i \equiv J$. 
It should be noted that the right-hand side of (\ref{3.1}) is the leading
term of an asymptotic expansion; its character changes from 
purely real (for $J^2 t^2 < n^2$) to complex (for $J^2 t^2 >
n^2$). More explicit forms are eq. (1.23) in ref. \onlinecite{VT78}
and eqs. (59,61) in ref. \onlinecite{MPS83b}.

We have calculated $ \langle
S_{N/2}^x(t) S_{N/2}^x \rangle$ for impurity coupling constants $0.1 \leq
J^{\prime} \leq 4$. In all cases the asymptotic decay of the correlation
function was consistent with the $t^{-1/2}$ law (\ref{3.1}). With growing
$J^{\prime}$ $| \langle S_{N/2}^x(t) S_{N/2}^x \rangle|$ develops oscillations
of rather well-defined frequency and growing amplitude, as shown in the inset
of Fig. \ref{eins}. The frequency of the oscillations for $J^{\prime} > 1$ is
proportional to the energy
\begin{equation}
  \label{eps0}
  |\varepsilon_0| = \frac{J^{\prime 2}}{\sqrt{2 J^{\prime 2}-1}} \quad
  (J^{\prime} > 1)
\end{equation}
of the localized impurity state.  On the whole, the long-time asymptotic
behavior of the impurity spin $x$ autocorrelation at $T=0$ is not
fundamentally changed by varying the value of $J^{\prime}$.
%%%%%%%%%%%%%%%%%%%%%%%%%%%%%%%%%%%%%%%%%%%%%%%%%%%%%%%%%%%%%%%%%%%%
\begin{figure}[h] 
\centerline{ \epsfxsize=9cm \epsfbox{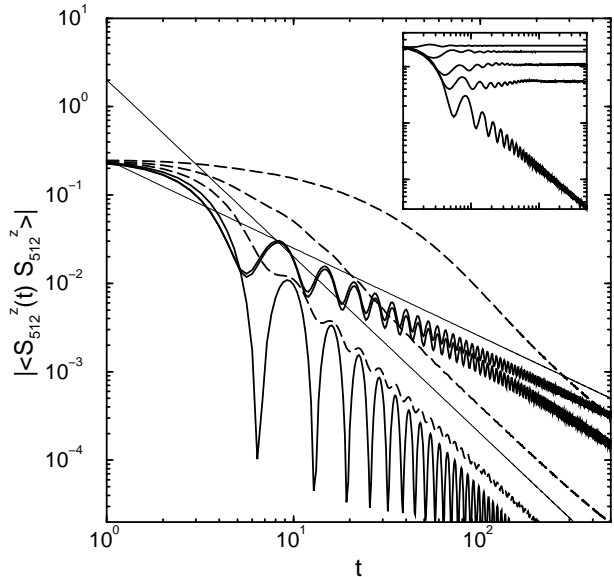} } 
\caption{Main plot:Bulk impurity spin $z$ autocorrelation function
  $|\langle S_{N/2}^z(t) S_{N/2}^z \rangle|$ $(N=1024)$ at $T=0$, for impurity
  couplings $J^{\prime}=1, 0.99, 0.8$ (solid lines, top to bottom);
  $J^{\prime}= 0.6, 0.4, 0.2$ (long-dashed lines, bottom to top). The thin
  straight lines show the power laws $t^{-1}$ and $t^{-2}$. Inset: same as
  main plot, for $J^{\prime}=1, 1.2, 1.4, 2, 4$ (bottom to top).}
\label{zwei} 
\end{figure}
%%%%%%%%%%%%%%%%%%%%%%%%%%%%%%%%%%%%%%%%%%%%%%%%%%%%%%%%%%%%%%%
The impurity spin $z$ autocorrelation, in contrast, changes significantly when
$J^{\prime}$ is varied, as shown in Fig. \ref{zwei}. For small $J^{\prime}$
$|\langle S_{N/2}^z(t)S_{N/2}^z\rangle|$ displays a monotonic decay. Roughly
at $J^{\prime}=0.5$ oscillations (of well-defined and $J^{\prime}$-independent
frequency) begin to develop. For all $J^{\prime}<1$ the correlation function
follows an asymptotic $t^{-2}$ law.  The $J^{\prime}=1$ correlation shows the
$t^{-1}$ law well known \cite{Nie67,KHS70,MS84} for the homogeneous case.  As
soon as $J^{\prime}$ is further increased, the behavior changes again and the
absolute value of the correlation function tends to a constant nonzero value
for large times.

The changes in the asymptotics of $|\langle S_{N/2}^z(t) S_{N/2}^z \rangle|$
can be understood from the analytic solution mentioned in Sec. \ref{II}. The
Jordan-Wigner transformation yields, after a few simple steps,
\begin{eqnarray}
  \label{3.2}
& &  \langle S_i^z(t) S_i^z\rangle = 
  \left(
    \sum_{\nu} |\langle i | \nu \rangle|^2 e^{i \varepsilon_{\nu} t}
  f(\varepsilon_{\nu}) 
  \right)^2\\ \nonumber
& &  +
  \left(
    \sum_{\nu} |\langle i | \nu \rangle|^2 
  f(\varepsilon_{\nu}) 
  \right)^2
  -
    \sum_{\nu} |\langle i | \nu \rangle|^2 f(\varepsilon_{\nu}) + \frac{1}{4}.
\end{eqnarray}     
Here $|\nu\rangle$ is a one-fermion eigenstate of $\tilde{H}$ (\ref{2.2}) with
energy $\varepsilon_{\nu}$ and $f(x)=(\exp(\beta x) +1)^{-1}$ is the Fermi
function. For $J^{\prime} > J_c =1$ the presence of a localized impurity state
with large $|\langle i | \nu \rangle|^2$ and with $\varepsilon_{\nu}$ outside
the continuum yields a harmonically oscillating non-decaying contribution to
$\langle S_i^z(t) S_i^z\rangle$.  Similar contributions are contained in every
element of the Pfaffian for $\langle S_i^x(t) S_i^x\rangle$, but not in that
correlation itself (see Fig. \ref{eins}, inset). The reason probably is a
cancellation of terms due to the multiplications and additions inherent in the
definition of the Pfaffian.  The time scale introduced by the discrete energy
value (\ref{eps0}) is reflected in the oscillations of $ | \langle
S_{N/2}^x(t) S_{N/2}^x \rangle| $ (Fig. \ref{eins}, inset). Similar behavior
is found for $T>0$ and will be discussed in the next subsection.

For $J^{\prime} \leq 1$ the time-dependent term in (\ref{3.2}) is proportional
to 
\begin{equation}
  \label{3.3}
  \left(
\int_{-1}^{1} d \varepsilon (1-\varepsilon^2)^{-\frac{1}{2}} e^{i \varepsilon
  t} f(\varepsilon) |\langle i|\varepsilon\rangle|^2
  \right)^2,
\end{equation}
where $\langle i|\varepsilon\rangle$ corresponds to $\langle i|\nu \rangle$ in
(\ref{3.2}) and the inverse square root factor is the one-particle density of
states of the dispersion (\ref{2.5}) (which still describes the energy
eigenvalues, only with slightly displaced $k$ values for $J^{\prime} \neq 1$).
For $J^{\prime}=1$, $|\langle i|\varepsilon\rangle|^2$ does not depend on
$\varepsilon$, the inverse square root singularities at the band edges
$\varepsilon \rightarrow \pm 1$ lead to a $t^{-1/2}$ asymptotic behavior of
the integral, and to a $t^{-1}$ behavior of $\langle S_i^z(t) S_i^z \rangle$.
For $J^{\prime}<1$ the amplitude of the one-particle eigenstate with energy
$\varepsilon$ at the impurity site is
\begin{equation}
  \label{3.4}
  |\langle i|\varepsilon\rangle|^2 =
  \left[
    J^{\prime 2} + \frac{\left( 1 - J^{\prime 2}\right)^2}{J^{\prime 2}}    
    \frac{\varepsilon^2 }{1 - \varepsilon^2}
  \right]^{-1}
\end{equation}
(apart from weakly $\varepsilon$-dependent normalization factors). This
changes the band-edge singularities in (\ref{3.3}) from $(1 -
\varepsilon^2)^{-1/2}$ to $(1 - \varepsilon^2)^{1/2}$, so that the integral
contains a $t^{-3/2}$ term. Consequently, $\langle S_i^z(t) S_i^z \rangle$
contains a $t^{-3}$ term which dominates for $T>0$ (see next subsection). At
$T=0$, however, the leading term is proportional to $t^{-2}$ due to the
discontinuity of the Fermi function in (\ref{3.3}).

We have also studied the $x$ and $z$ autocorrelations of nearest and
next-nearest neighbors of the impurity spin $i=N/2$. For weak impurity
coupling ($J^{\prime} \lesssim 0.2$) the $x$ correlation functions of spins
$i=N/2+1$ and $i=N/2+2$ show weak oscillations superimposed on a $t^{-1/2}$
decay masked by strong finite-size effects. The $z$ correlations (for
$J^{\prime}<1$) show stronger oscillations. Their decay looks roughly like a
power law with an exponent between -2 and -3.

\subsection{$T>0$}
\label{IIIB}
%%%%%%%%%%%%%%%%%%%%%%%%%%%%%%%%%%%%%%%%%%%%%%%%%%%%%%%%%%%%%%%%%%%%%%%%%
\begin{figure}[h] 
\centerline{ \epsfxsize=9cm \epsfbox{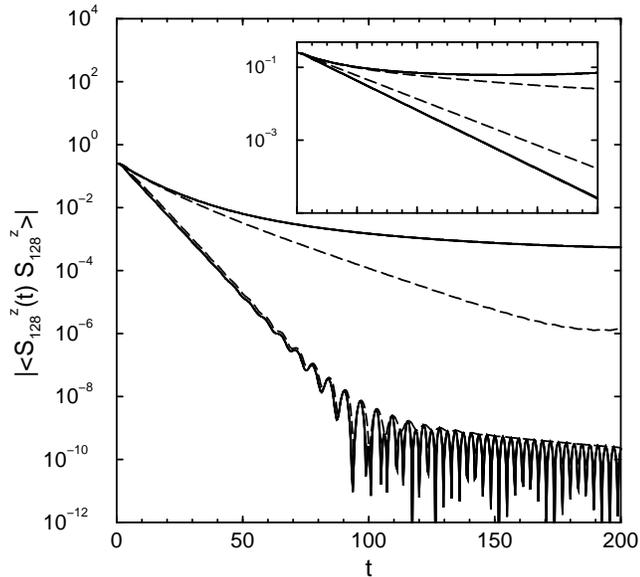} } 
\caption{Crossover between the regimes of  high temperature (lower set of
  curves) and low temperature (upper set of curves). Main plot: Impurity spin
  autocorrelation function  $| \langle S_{N/2}^z(t) S_{N/2}^z \rangle | $
  ($N=256$) with impurity coupling  $J^{\prime}=0.3$ for temperatures 
  $T=10^6$, $10^4$, $10^2$, $10^0$, $10^{-1}$, $10^{-2}$, $10^{-3}$,
  $10^{-4}$, $10^{-5}$,  and $10^{-6}$ (bottom to top). The curves for
  $T=10^{-1}$ and $10^{-2}$ are 
  long-dashed, all others are solid. Note that several
 curves coincide on the scale of
  the figure. Inset: same as main plot, for
  $| \langle S_{N/2}^x(t) S_{N/2}^x \rangle | $  ($N=128$); time range is 
  $0 < t <100$.}
\label{drei} 
\end{figure}
%%%%%%%%%%%%%%%%%%%%%%%%%%%%%%%%%%%%%%%%%%%%%%%%%%%%%%%%%%%%%%%%%%%%%%%%%
Whereas at $T=0$ only power-law decay is observed, exponential decay
\cite{exdec}  becomes
possible at finite $T$.  Fig. \ref{drei} shows $x$ and $z$ impurity spin
autocorrelations at $J^{\prime}=0.3$ for several decades in $T$. There are two
well defined temperature regimes with a crossover between them. Within each
regime the correlation functions do not change qualitatively: note that
several of the curves in Fig. \ref{drei} coincide.  The $x$ autocorrelation in
the high-$T$ regime shows exponential decay which persists over the entire
time range during which the the results are free of finite-size effects. The
$z$ autocorrelation decays exponentially at first and later crosses over to
the $t^{-3}$ law derived above, with superimposed oscillations which are
absent in the low-temperature regime.  
The appearance of oscillations
(if only of small amplitude) in $\langle S_i^z(t) S_i^z \rangle$ at high $T$
is reminiscent of the phenomena recently reported \cite{SH98} for a two-level
system coupled to a spin bath. In that system, a persistence of oscillations
up to infinite $T$ could be observed.
%%%%%%%%%%%%%%%%%%%%%%%%%%%%%%%%%%%%%%%%%%%%%%%%%%%%%%%%%%%%%%%%%%%%%%%%%
\begin{figure}[h] 
\centerline{ \epsfxsize=9cm \epsfbox{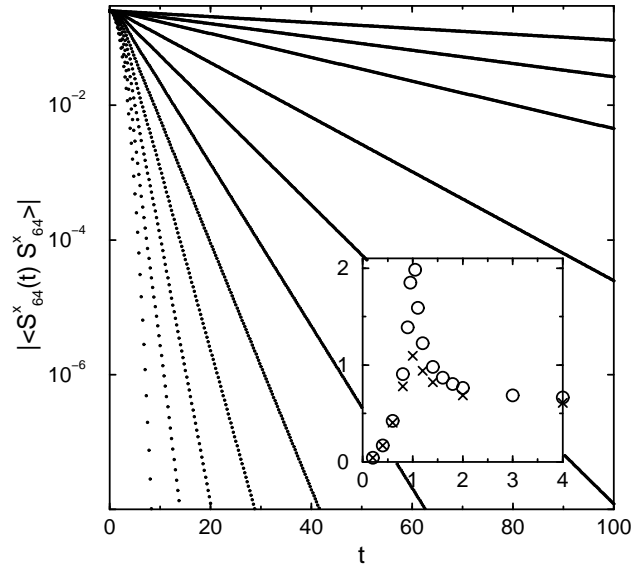} } 
\caption{Exponential decay of the bulk impurity spin $x$ autocorrelation
  function at high temperature.
 Main plot: $|\langle S_{N/2}^x(t) S_{N/2}^x \rangle|$ $(N=128)$ at $T=10^5$
 for impurity couplings $J^{\prime}=0.1, 0.15, 0.2, 0.3, 0.4, 0.5, 0.6, 0.7,
 0.8, 0.9,$ and 1 (top to bottom).
  Inset: Exponential decay rates determined from the data shown in the main
 plot, and from analogous data for $J^{\prime}>1$ and for $T=1$. The circles
  are $T=10^5$ data, the crosses are $T=1$ data. Note the pronounced
 differences near $J^{\prime}=1$, where the infinite-$T$ decay rate diverges.}
\label{vier} 
\end{figure}
%%%%%%%%%%%%%%%%%%%%%%%%%%%%%%%%%%%%%%%%%%%%%%%%%%%%%%%%%%%%%%%%%%%%%%%%%
Fig. \ref{vier} shows $| \langle S_{N/2}^x(t) S_{N/2}^x \rangle | $ at
$T=10^5$ for impurity coupling $0.1 \leq J^{\prime} \leq1$ in a $N=128$ chain.
As autocorrelations at $T=\infty$ are real even power series in $t$, all
curves start with zero slope at $t=0$, but then (with the exception of
$J^{\prime}=1$) bend over to a nearly perfect exponential decay. The inset
shows the decay rate of that exponential decay as fitted to the data in the
main plot. Also shown is the decay rate determined from data for $T=1$, and
for $J^{\prime} > 1$.  Differences between $T=1$ and $T=\infty$ are to be
expected and are indeed visible in the behavior of the decay rate close to
$J^{\prime}=1$: For finite $T$ the $x$ autocorrelation function of a
homogeneous chain $J_i\equiv1$ is known \cite{IIKS93,my28} to decay
exponentially with a finite $T$-dependent decay rate, whereas for infinite $T$
the decay is Gaussian \cite{BJ76,CP77,FL87}. The Gaussian decay
corresponds to a 
divergence of the exponential decay rate which is obvious in the inset of Fig.
\ref{vier}.

For $J^{\prime}>1$ $| \langle S_{N/2}^x(t) S_{N/2}^x \rangle | $ is no longer
(almost) purely exponential, but develops considerable
oscillations. The exponential decay rate grows with $T$, but decreases as
$J^{\prime}$ grows, as shown in the inset of Fig. \ref{vier}. As in the  $T=0$
case the frequency of the oscillations is proportional to $\varepsilon_0$
 (\ref{eps0}).

In order to obtain a quantitative measure of the precision to which the decay
of $| \langle S_{N/2}^x(t) S_{N/2}^x \rangle | $ at $T=10^5$
follows an exponential, we fitted an exponential law $a \exp (-bt)$ to the
numerical data for $ 0 < t < 100$ and calculated the quantity
\begin{equation}
  \label{prec}
  p(t) := \frac{| \langle S_{N/2}^x(t) S_{N/2}^x \rangle |}{a \exp (-bt)}
\end{equation}
which should equal unity for a purely exponential decay. 
%%%%%%%%%%%%%%%%%%%%%%%%%%%%%%%%%%%%%%%%%%%%%%%%%%%%%%%%%%%%%%%%%%%%%%%%%
\begin{figure}[h] 
\centerline{ \epsfxsize=9cm \epsfbox{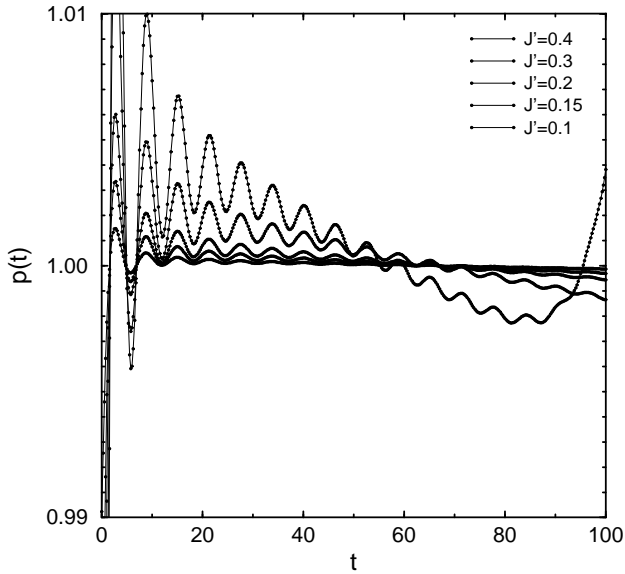} } 
\caption{ Precision of the exponential decay of the high-temperature bulk
  impurity spin correlation function. Shown is the precision function $p(t)$
  (\ref{prec}) for the data for $J^{\prime} \leq 0.4$ from the main plot of
  Fig. \ref{vier}. }
\label{vierb} 
\end{figure}
%%%%%%%%%%%%%%%%%%%%%%%%%%%%%%%%%%%%%%%%%%%%%%%%%%%%%%%%%%%%%%%%%%%%%%%%%
For $J^{\prime} \leq 0.4$, $p(t)$ is shown in Fig. \ref{vierb}. Note that the
scale of the figure extends only to a maximum deviation of 1 percent from
purely exponential decay. The general slope in the data is a natural
consequence of the intrinsically non-exponential behavior of the correlation
function for small $t$ which was mentioned above.  The data for $J^{\prime}
\leq 0.2$ follow the purely exponential fit to a precision of better than 2
parts in thousand for $t \geq 10$. This rules out the stretched-exponential
behavior reported \cite{SB98} for $J^{\prime} \leq 0.2$ at $T=\infty$ in an
approximate study based on extrapolation of truncated continued-fraction
expansions. Fig. \ref{vierb} also reveals the presence of tiny oscillations
which are invisible on the scale of Fig. \ref{vier}. The frequency of these
oscillations is independent of $J^{\prime}$, in contrast to the
stronger oscillations for $J^{\prime} > 1$ already mentioned above.
%%%%%%%%%%%%%%%%%%%%%%%%%%%%%%%%%%%%%%%%%%%%%%%%%%%%%%%%%%%%%%%%%%%%%%%%%
\begin{figure}[h] 
\centerline{ \epsfxsize=9cm \epsfbox{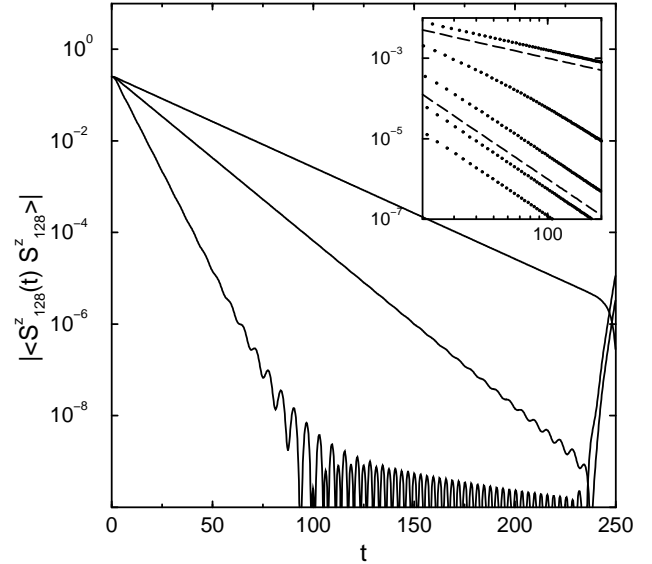} } 
\caption{Initially exponential and asymptotically power-law decay of the bulk
  impurity $z$ spin autocorrelation function at high temperature.  Main plot: 
$| \langle S_{N/2}^z(t) S_{N/2}^z \rangle |$ ($N=256$) at $T=10^5$ for impurity
couplings $J^{\prime}=0.15, 0.2,$ and 0.3 (top to bottom). Note how the
exponential regime gets shorter as $J^{\prime}$ grows. The features in the
lower right corner are finite-size effects.  
  Inset: Maxima of $| \langle S_{N/2}^z(t) S_{N/2}^z \rangle |$ for
  $J^{\prime} = 1, 0.9, 0.8, 0.7,$ and 0.6 (top to bottom) and $T \geq
  20$. The dashed straight lines represent the $t^{-1}$ and $t^{-3}$ power
  laws.} 
\label{fuenf} 
\end{figure}
%%%%%%%%%%%%%%%%%%%%%%%%%%%%%%%%%%%%%%%%%%%%%%%%%%%%%%%%%%%%%%%%%%%%%%%%%
The main differences between $x$ and $z$ spin autocorrelation functions were
already shown in Fig. \ref{drei}. Some more detail on the behavior of $|
\langle S_{N/2}^z(t) S_{N/2}^z \rangle | $ is presented in Fig. \ref{fuenf}.
The main plot shows the crossover between the exponential and $t^{-3}$ (with
superimposed oscillations) regimes for three small $J^{\prime}$ values. With
growing $J^{\prime}$ the exponential decay rate grows and the exponential
regime shortens in such a way that in the exponential regime $| \langle
S_{N/2}^z(t) S_{N/2}^z \rangle | $ is a decreasing function of $J^{\prime}$
whereas in the power-law regime it is an increasing function of $J^{\prime}$.
We have deliberately chosen a very long time window in order to illustrate how
finite-size effects manifest themselves (for $ t \gtrsim 230$).  In the inset
of Fig. \ref{fuenf} we demonstrate the asymptotic power-law behavior for
larger values of $J^{\prime}$. The dots represent the maxima of $| \langle
S_{N/2}^z(t) S_{N/2}^z \rangle | $, which follow the $t^{-3}$ (for
$J^{\prime}<1$) and $t^{-1}$ (for $J^{\prime}=1$) laws already discussed.  For
$J^{\prime}>1$ the behavior changes to a ``constant with oscillations'' type
of asymptotics, similar to the $T=0$ situation shown in the inset of Fig.
\ref{zwei}. For $T>0$, however, the amplitude of the oscillations is
considerably larger than for $T=0$.

The high-temperature spin autocorrelations of the nearest and next-nearest
neighbors to the impurity (for $J^{\prime}\leq1$) do not show any particularly
surprising features. The $z$ correlations do not show exponential decay in the
beginning. Instead they oscillate and the maxima of the oscillations display
the familiar $t^{-3}$ (for $J^{\prime}<1$) and $t^{-1}$ (for $J^{\prime}=1$)
laws. The $x$ autocorrelations interpolate smoothly between two known limiting
cases. At $J^{\prime}=1$ the $x$ autocorrelation of the spin $i=N/2+1$ of
course is a Gaussian as that of any other bulk spin. At $J^{\prime}=0$,
however, $i=N/2+1$ is the first spin in a semi-infinite homogeneous chain,
whose $T=\infty$ autocorrelation function is \cite{my18} a combination of
Bessel functions with an asymptotic $t^{-3/2}$ decay. Upon reducing
$J^{\prime}$ from 1 to 0, the development of the characteristic Bessel
function oscillations (with zeros hardly depending on $J^{\prime}$) can be
nicely observed. Similarly the time range during which the correlation
function follows the expected $t^{-3/2}$ decay grows as $J^{\prime}$
diminishes. The $x$ autocorrelation of $i=N/2+2$ behaves quite similarly, but
only a small number of oscillations is visible (in a linear plot) because of
the fast ($t^{-9/2}$) asymptotic decay of the known \cite{my18} $J^{\prime}=0$
Bessel function expression.
\section{Boundary impurity}
\label{IV}
The boundary impurity is defined by (\ref{2.7}).
Similarly to the case of the bulk impurity discussed in the previous section,
the boundary correlation functions show a low-temperature regime and a
high-temperature regime, and we  restrict our discussion to the values
$T=0$ and $T=1$ which represent these two regimes. 

It suffices to discuss the impurity spin $x$ autocorrelation function 
$ \langle S_1^x(t) S_1^x \rangle $, because
\begin{equation}
  \label{4.1}
  2 \langle S_1^x(t) S_1^x \rangle = \sum_{\nu} |\langle 1 | \nu \rangle |^2
  e^{i \varepsilon_{\nu} t} f(\varepsilon_{\nu}),  
\end{equation}
that is, the square root of the time-dependent part of $\langle S_1^z(t) S_1^z
\rangle$ (see \ref{3.2}).  The presence of an isolated impurity state for
$J^{\prime} > J_c = \sqrt{2}$ (compare Sec. \ref{II}) should be visible in the
dynamic correlation functions. In fact, the asymptotic behavior of $ \langle
S_1^x(t) S_1^x \rangle $ displays a crossover similar to the one shown in the
inset of Fig. \ref{zwei}: for all $T$ the long-time behavior is a power law
for $J^{\prime} < \sqrt{2}$ and a constant for $J^{\prime} > \sqrt{2}$, with
additional oscillations in both regimes.
%%%%%%%%%%%%%%%%%%%%%%%%%%%%%%%%%%%%%%%%%%%%%%%%%%%%%%%%%%%%%%%%%%%%%%%%%
\begin{figure}[h] 
\centerline{ \epsfxsize=9cm \epsfbox{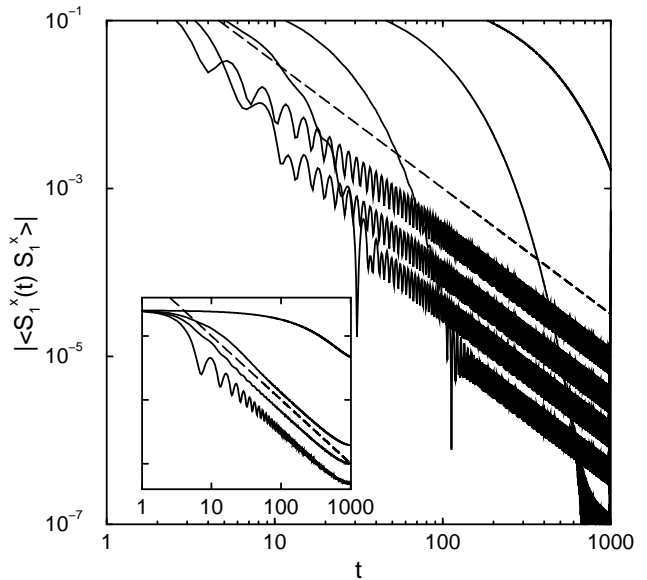} } 
\caption{Boundary spin $x$ autocorrelation function 
$| \langle S_1^x(t) S_1^x \rangle|$.
 Main plot: High-temperature regime, $T=1$, $N=512$, and impurity coupling 
$J^{\prime} =1$, 0.8, 0.6, 0.4, 0.2, and 0.1 (left to right, according to the
 point where the curves enter the figure at the upper boundary). The dashed
 line shows the $t^{-3/2}$ law.
  Inset: Low-temperature regime, $T=0$, $N=1024$, and $J^{\prime} =1$, 0.7,
 0.5, and 0.1 (bottom to top). The dashed line shows the $t^{-1}$ law.}
\label{sechs} 
\end{figure}
%%%%%%%%%%%%%%%%%%%%%%%%%%%%%%%%%%%%%%%%%%%%%%%%%%%%%%%%%%%%%%%%%%%%%%%%%
Fig. \ref{sechs} shows $| \langle S_1^x(t) S_1^x \rangle|$ for impurity
coupling $J^{\prime} \leq 1$. For $T=1$ we observe an initially exponential
decay followed by a power law. The exponential decay rate grows with
$J^{\prime}$. The duration of the initial exponential phase decreases with
growing $J^{\prime}$ in such a way that in the subsequent power-law regime the
correlation function is an increasing function of $J^{\prime}$. This behavior
is (not unexpectedly, compare (\ref{4.1})) similar to that of the bulk impurity $z$ autocorrelation
discussed in Sec. \ref{IIIB}, however, the asymptotic power law is a different
one, namely the $t^{-3/2}$ law known for the boundary spin \cite{my18} of the
homogeneous semi-infinite chain at infinite $T$, and for a range of boundary
spins \cite{my28} of the same system at finite $T$.

At this point, the early analytical study by Tjon \cite{Tjo70} should be
mentioned. In the limit of sufficiently small $J^{\prime}$ Tjon found an
asymptotically exponential decay $\sim e^{-t/\tau}$ for $ \langle S_1^x(t)
S_1^x \rangle $, with decay rate 
$\tau^{-1} = \frac{1}{2} J^{\prime 2}$. Indeed, our numerical data show that
during the exponential 
regime mentioned above, the decay rate is quite precisely equal to 
$\frac{1}{2} J^{\prime 2}$ for $J^{\prime} \lesssim 0.2$, and a bit larger for
larger $J^{\prime}$. However, we also observe numerically (and explain
analytically, see below) a crossover from exponential to power-law
behavior. The duration of the exponential regime grows ($\sim J^{\prime -2}
$) as $J^{\prime}$ becomes weak and thus our numerical results for arbitrary
$J^{\prime}$ connect smoothly to Tjon's analytical result restricted to the
weak-coupling limit.

The inset of Fig. \ref{sechs} shows $| \langle S_1^x(t) S_1^x \rangle |$ for
$T=0$ and $J^{\prime} <1$. After a very slow initial decay (slower for smaller
$J^{\prime}$) the curves eventually all bend over to show a $t^{-1}$
asymptotic decay developing some oscillations as $J^{\prime}$ approaches
unity. The $t^{-1}$ decay at $T=0$ is again a well-known feature \cite{PM78} of
the homogeneous semi-infinite chain.

The asymptotic power laws may be understood from the properties of the
analytic solution of the boundary impurity (one-particle) problem, along the
lines of the discussion in Sec. \ref{IIIA}. According to (\ref{4.1}), $|
\langle S_1^x(t) S_1^x \rangle |$ for $J^{\prime} < \sqrt{2}$ is given by an
integral analogous to the one in (\ref{3.3}).  The analytic solution shows
that the wave function factor in the integral, $ \langle 1 | \varepsilon
\rangle \sim \sin k$, where $\varepsilon = - \cos k$. This leads to a
band-edge singularity $ \sim (1-\varepsilon)^{1/2}$ in the integrand, and a
$t^{-3/2}$ asymptotic behavior of the integral. At $T=0$ the $t^{-3/2}$
contribution is dominated by the $t^{-1}$ contribution from the discontinuity
of the Fermi function.

\section{Summary and conclusions}
\label{V}

The dynamic spin correlation functions associated with isolated impurities in
a $S=1/2$ $XX$ chain show a rich behavior depending on the temperature, the
impurity coupling strength $J^{\prime}$, the spin component ($\alpha=x,z$)
under consideration, and on the position of the impurity spin in the chain.
Regimes of low and high $T$, with qualitatively different behavior of the
correlations may be distinguished.  We have summarized the asymptotic behavior
of the bulk $x$ and $z$ spin autocorrelations in these two $T$ regimes in
Table \ref{un}. The basic features of the boundary correlations may also be
obtained from Table \ref{un} by observing that (i) the $z$ autocorrelation
does not change fundamentally between bulk and boundary and (ii) at the
boundary the $x$ autocorrelation behaves as the square root of the $z$
autocorrelation.

As mentioned in the introduction, the present model may be interpreted as a
single two-level system in contact with a large bath. The influence of the
nature of the bath degrees of freedom on the type of decay of the two-level
system was addressed recently \cite{SH98} for a spin bath constructed in a way
to resemble closely the standard \cite{Wei93} harmonic oscillator bath. The
following results were found. At $T=0$ the spin bath leads to damped
oscillations in the two-level system, as does the oscillator bath. However, at
high $T$, the oscillations vanish for the oscillator bath but persist for the
spin bath.

Without going into any detailed comparison between the spin bath employed in
Ref. \onlinecite{SH98} and the $XX$ chain studied here, we would like to point
out the existence of similar oscillation phenomena at high $T$ in the present
system. Fig. \ref{drei} (main plot) shows how the bulk impurity spin $z$
autocorrelation function develops oscillations as temperature {\em increases}.
Whether these oscillations can be unambiguously assigned to either the
impurity or the bath, and what happens for systems interpolating between the
present one and that of Ref.\onlinecite{SH98}, remains to be seen in further
studies.
%%%%%%%%%%%%%%%%%%%%%%%%%%%%%%%%%%%%%%%%%%%%%%%%%%%%%%%%%%%%%%%%%%%%%%%%%
\begin{table}[b] 
  \begin{tabular}{ccc}
 
 & low $T$ & high $T$\\
\hline\\
$x$ & $t^{-1/2}$  & $e^{-\frac{t}{\tau}}$  \\
\hline\\
$z, J^{\prime}<1$ & $t^{-2}$    & $t^{-3}$ \\
                  &             & (after initial $e^{-\frac{t}{\tau}}$) \\
\hline\\
$z, J^{\prime}>1$ & $t^{0}$    & $t^{0}$ \\

  \end{tabular}

\caption{Asymptotic decay of the bulk $x$ and $z$ autocorrelation
  functions. Additional oscillations of varying strength are present in all
  cases.} 
\label{un} 
\end{table}
%%%%%%%%%%%%%%%%%%%%%%%%%%%%%%%%%%%%%%%%%%%%%%%%%%%%%%%%%%%%%%%%%%%%%%%%%
\acknowledgments
We are grateful to Professor Gerhard M\"uller (University of Rhode Island) for
helpful comments and suggestions.

\newcommand{\noopsort}[1]{}


\begin{thebibliography}{10}

\bibitem{LSM61}
E. Lieb, T. Schultz, and D. Mattis, Ann. Phys. {\bf 16},  407  (1961).

\bibitem{Kat62}
S. Katsura, Phys. Rev. {\bf 127},  1508  (1962).

\bibitem{Nie67}
T. Niemeijer, Physica {\bf 36},  377  (1967).

\bibitem{KHS70}
S. Katsura, T. Horiguchi, and M. Suzuki, Physica {\bf 46},  67  (1970).

\bibitem{Tjo70}
J.~A. Tjon, Phys.\ Rev.\ B {\bf 2},  2411  (1970).

\bibitem{MBA71}
B.~M. McCoy, E. Barouch, and D.~B. Abraham, Phys.\ Rev.\ A {\bf 4},  2331
  (1971).

\bibitem{SJL75}
A. Sur, D. Jasnow, and I.~J. Lowe, Phys.\ Rev.\ B {\bf 12},  3845  (1975).

\bibitem{BJ76}
U. Brandt and K. Jacoby, Z. Phys. B {\bf 25},  181  (1976).

\bibitem{CP77}
H.~W. Capel and J.~H.~H. Perk, Physica {\bf 87A},  211  (1977).

\bibitem{PM78}
W. Pesch and H.~J. Mikeska, Z. Phys. B {\bf 30},  177  (1978).

\bibitem{VT78}
H.~G. Vaidya and C.~A. Tracy, Physica {\bf 92A},  1  (1978).

\bibitem{GC80}
L.~L. Gon\c{c}alves and H.~B. Cruz, J. Magn. Magn. Mat. {\bf 15-18},  1067
  (1980).

\bibitem{CG81}
H.~B. Cruz and L.~L. Gon\c{c}alves, J. Phys. C {\bf 14},  2785  (1981).

\bibitem{MPS83a}
B.~M. McCoy, J.~H.~H. Perk, and R.~E. Shrock, Nucl. Phys. {\bf B220 [FS8]},  35
   (1983).

\bibitem{MPS83b}
B.~M. McCoy, J.~H.~H. Perk, and R.~E. Shrock, Nucl. Phys. {\bf B220 [FS8]},
  269  (1983).

\bibitem{MS84}
G. M\"uller and R.~E. Shrock, Phys.\ Rev.\ B {\bf 29},  288  (1984).

\bibitem{TM85}
J.~H. Taylor and G. M\"uller, Physica {\bf 130A},  1  (1985).

\bibitem{FL87}
J. Florencio and M.~H. Lee, Phys.\ Rev.\ B {\bf 35},  1835  (1987).

\bibitem{my18}
J. Stolze, V. Viswanath, and G. M\"uller, Z. Physik B {\bf 89},  45  (1992).

\bibitem{IIKS93}
A.~R. Its, A.~G. Izergin, V.~E. Korepin, and N.~A. Slavnov, Phys.\ Rev.\ Lett.
  {\bf 70},  1704  (1993).

\bibitem{my28}
J. Stolze, A. N\"oppert, and G. M\"uller, Phys. Rev. B {\bf 52},  4319  (1995).

\bibitem{LCDFGZ87}
A.~J. Leggett {\it et~al.}, Rev.\ Mod.\ Phys. {\bf 59},  1  (1987).

\bibitem{Wei93}
U. Weiss, {\em Quantum dissipative systems} (World Scientific, Singapore,
  1993).

\bibitem{HTB90}
P. H\"anggi, P. Talkner, and M. Borkovec, Rev.\ Mod.\ Phys. {\bf 62},  251
  (1990).

\bibitem{SH98}
J. Shao and P. H\"anggi, Phys.\ Rev.\ Lett. {\bf 81},  5710  (1998).

\bibitem{PTVF92}
W.~H. Press, S.~A. Teukolsky, W.~T. Vetterling, and B.~P. Flannery, {\em
  Numerical Recipes in FORTRAN: the Art of Scientific Computing} (Cambridge
  U.P., Cambridge, 1992).

\bibitem{Pfaffnote}
The properties of Pfaffians, close relatives of determinants, are discussed in
  ref. \onlinecite{GH64}, the explicit formula for $\langle S_i^{x}(t) S_j^{x}
  \rangle$ is given in ref. \onlinecite {my28}.

\bibitem{note2}
For the bulk impurity (\ref{2.8}) a system with $N-1$ sites should be
  considered for symmetry.

\bibitem{exdec}
The real (imaginary) part of the autocorrelation function of a hermitian
  operator is an even (odd) function of time, hence purely exponential decay is
  not possible. A theorem ruling out purely exponential decay for a broad class
  of systems was stated by Lee \cite{Lee83}. In the present paper, we always
  refer to an (approximately) exponential decay in a time range limited for
  general reasons at short times and by finite-size effects at long times.

\bibitem{SB98}
S. Sen and T.~D. Blersch, Physica A {\bf 253},  178  (1998).

\bibitem{GH64}
H.~S. Green and C.~A. Hurst, {\em Order-Disorder Phenomena}
  (Wiley-Interscience, London, 1964).

\bibitem{Lee83}
M.~H. Lee, Phys.\ Rev.\ Lett. {\bf 51},  1227  (1983).

\end{thebibliography}
\end{document}